\documentstyle[12pt]{article}
\title{Exact Solutions of 1+1 Dimensional Dilaton Gravity
Coupled to Matter}
\author{A.T. Filippov\\
{\it Joint Institute for Nuclear Research, Dubna, Russia} \\
{\it RIMS, Kyoto University, Kyoto, Japan} \\
{\small e-mail: filippov@thsun1.jinr.dubn.su}}
\date{}  

\def\half{\frac{1}{2}}

\def\metl{g_{ij}}
\def\metlb{\bar{g}_{ij}}
\def\metu{g^{ij}}

\def\eml{F_{ij}}
\def\emu{F^{ij}}

\def\lf{(\log{f})_{uv}}
\def\Lh{\log{h}}
\def\Lhd{(\log{h}){\dot{ }}}
\def\sqd{){\dot{ }}}

\def\fid{\dot{\phi}}
\def\fids{{\dot{\phi}}^2}
\def\psid{\dot{\psi}}
\def\psids{{\dot{\psi}}^2}
\def\hd{\dot{h}}

\def\fidd{\ddot{\phi}}
\def\be{\begin{equation}}
\def\ee{\end{equation}}
\def\bea{\begin{eqnarray}}
\def\eea{\end{eqnarray}}
\def\tf{\tilde{f}}
\def\tV{\tilde{V}}

\def\dif{\partial}

\def\Hamb{{\bar{\cal H}}^{(s)}}

\def\ham{H^{(s)}}
\def\hamb{{\bar{H}}^{(s)}}

\def\s{\hspace{1mm}}
\def\EXP{{\rm e}}

\begin{document}

\maketitle

\begin{center}
\bf{Abstract}
\end{center}

{\small
\noindent
A class of integrable models of 1+1 dimensional dilaton gravity
coupled to scalar and electromagnetic fields is obtained and explicitly
solved. More general models are reduced to 0+1 dimensional Hamiltonian
systems, for which two integrable classes (called s-integrable) are found
and explicitly solved. As a special case, static spherical solutions of the
Einstein gravity coupled to electromagnetic and scalar fields in any real
space-time dimension are derived. A generalization of the `no-hair'
theorem is pointed out and the Hamiltonian formulation that enables
quantizing s-integrable systems is outlined.
}

\bigskip
\noindent {\bf 1.}
The general 1+1 dimensional dilaton gravity \cite{Banks} recently
attracted a good deal of attention in connection with the black hole physics
(for a review and references see e.g. \cite{Mann}-\cite{Kunst}).
An important special case of this theory -- the 1+3
dimensional Schwarzschild black hole (SBH) -- was recently attempted to
quantize by using different approaches \cite{Kastrup}-\cite{VDA}.
The key property of SBH, which makes the
quantization possible, is integrability\footnote{
For a relevant review of integrable dynamical systems see e.g.
\cite{Andrei}, \cite{Asik}. Integrable systems considered here are
explicitly integrable, meaning that their classical solutions can be
found in terms of elementary functions or reduced to quadratures.}
\cite{Kastrup}, \cite{VDA}.                                        
In fact, the most general 1+1 dilaton gravity (DG) also has this nice property.
Due to the generalized Birkhoff theorem \cite{Kunst}, it actually reduces
to a finite dimensional constrained system (FDC) \cite{Banks}, and thus
can be explicitly quantized following the approach of Ref.\cite{VDA}.

Though quantizing SBH might seem to be most important, in view of their
special role in the Einstein gravity theory, the more general models
have to be considered. An explicitly integrable system related to
string models and not reducible to FDC had been proposed in
\cite{CGHS} (CGHS)
and was analyzed in detail for the last three years (see e.g. \cite{Strom},
\cite{Jackiw} and references therein). A more complex integrable
model, related to the
Kerr black hole, was recently studied with employing the full machinery of
modern methods for quantizing integrable systems (see \cite{Nicolai} where
references to other papers of the authors may be found).

Though such models are integrable, explicit quantum solutions of them
are not easy to construct. For this reason, quantizing reduced models
corresponding to the static spherically symmetric sector of gravity coupled
to scalars is of significant importance \cite{Marco}. One of the consequences
of the famous `no hair' theorems (for a general formulation and references
see \cite{Beken}) is that the scalar energy - momentum tensor can not be
treated as a perturbation on the SBH background. However, one may hope that
a perturbative treatment of non static solutions around exact static
ones may still be viable.

All this motivates our search for DG models coupled to scalars (DGS)
which are completely integrable or, at least, integrable in the
`static sector' (the precise meaning of this term will be defined later,
we will call them s-integrable).
We first formulate a general DGS theory, briefly review its relation
to the gravity theory in any space - time dimension, and present a very
simple solution of the most general DG. Our approach is then applied to
constructing apparently new integrable and s-integrable models.
The Hamiltonian formulation of the 0+1 dimensional models is also outlined.
In the s-integrable case this allows one to apply our approach \cite{VDA}
to quantization.  As a
byproduct, we point out certain generalization of the `no hair' theorem.
Our treatment is completely local (not using boundary conditions,
causality, etc.) but the derived exact solutions may provide a good
starting point for a global analysis.

\medskip
\noindent {\bf 2.}
The general 1+1 dimensional DG treated here\footnote{
Many authors consider the models with $W = {\rm const}$ (usually
$\half$ or even $0$, see e.g. \cite{Banks}, \cite{Strobl}, \cite{Kunst},
\cite{Borya}). As will soon become clear, it is important
to keep $W$ arbitrary. Note that we call the dilaton gravity (DG)
the integrable $U-X$ part of the Lagrangian (1).
The general $U-Z$ model, which we call DGS, may be extended to include
many scalar fields, nonabelian gauge fields and spinor fields.
Such models naturally emerge in the dimensionally reduced Einstein
gravity coupled to matter (see e.g. \cite{Haj}). }
is given by the Lagrangian
$$
{\cal L} = \sqrt{-g} \, [U R + V + W\metu \phi_i \phi_j + X\eml \emu + Y +
Z\metu \psi_i \psi_j] \, ,                                        \eqno(1)
$$
where $U,\;V,\;W,\;X,\;Y,\;Z$ are functions of the dilaton field $\phi$
while $Y$ may in addition depend on the scalar field $\psi$; $\metu$ is
the contravariant metric tensor, $R$ the scalar curvature, $\eml$ the
electromagnetic field tensor. In this paper,
the lower letter indices denote partial derivatives ($\phi_i =
\partial_i \phi$, etc.), except when used in
$\metl$ or $\eml$. The equations of motion may always be written in the
metric $ds^2 = -4f(u,v)du dv$. To simplify the first equation we define
the function $w(\phi)$, the importance of which will become clear in a
moment, $w^{\prime}/w \equiv W/U^{\prime}$ (the prime always denotes the
derivative of a function depending on one variable, thus $U^\prime \equiv
dU/d\phi$, etc.). Omitting the standard derivations (see e.g.
\cite{Wald}, \cite{Kunst}), we write all equations in a convenient
form ($Y_\phi \equiv \dif_{\phi} Y,\ Y_\psi \equiv \dif_{\psi} Y$)
$$
f w(U_i /f w)_i = Z \psi_i^2\;,\;\;(i = u,v) \, ;  \eqno(2)
$$
$$
U_{uv} + f(V - X\eml \emu + Y) = 0; \;\;           \eqno(3)
$$
$$
2W\phi_{uv} + W^{\prime}\phi_u \phi_v + U^{\prime} \lf +
f(V^{\prime} + X^{\prime}\eml \emu + Y_{\phi}) =
Z^{\prime} \psi_u \psi_v\ \, ;                     \eqno(4)
$$
$$
(X F_{uv}/2f)_i = 0,\;\;(i = u,v) \, ;             \eqno(5)
$$
$$
(Z\psi_u)_v + (Z\psi_v)_u + f Y_{\psi} = 0 \, .    \eqno(6)
$$
The equations (2) and (3) were derived by varying ${\cal L}$ in $\metu$,
the third term in eq.(4) is proportional to the curvature $R =
f^{-1}(\log{f})_{uv}$. Eq.(5) gives the first integral of motion
$Q = XF_{uv}/2f$, then $-X \eml \emu = 2Q^2/X$. It follows that
$-X \eml \emu $ may be included in the potential $V$.

An interesting class of DG described by (1) can be derived by a dimensional
reduction of the spherically symmetric sector of the $d$-dimensional
Einstein gravity
coupled to scalar and electromagnetic fields having the same symmetry.
Following Ref.\cite{Soda} (see also \cite{Uehara}), let us assume that
the $d$-dimensional metric has the form ($d \equiv n + 2$, $\nu \equiv 1/n$)
$$
G_{I J} = \pmatrix{
	    \metl    &0\cr
	    0        &\EXP^{-2\nu\phi} g_{{\bar{i}} {\bar{j}}}\cr},
$$
where $g_{{\bar{i}} {\bar{j}}}$ is the metric of the $n$-dimensional sphere
(more generally, a maximally symmetric compact space \cite{Soda}).
As the fields
$\phi, \psi, \metl, \eml$ depend only on the coordinates of the
1+1 dimensional subspace, we may integrate over the coordinates of
the sphere and get the effective Lagrangian in the form (1), where
$$
U = \EXP^{-2\phi},\;\;V = 2\EXP^{-2\phi}(\alpha \EXP^{4\nu \phi} + \Lambda
- \beta Q^2 \EXP^{4\phi}),\;\; W = 4(1-\nu)\EXP^{-2\phi},\; \; \eqno(7a)
$$
$$
 Y = -\gamma y(\psi)\EXP^{-2\phi},\;\;\;
 Z = -\gamma\EXP^{-2\phi}. \;\; \eqno(7b)
$$
Here the parameters $\alpha, \beta, \gamma$ may depend on $\nu$; $\alpha$ is
proportional to the curvature of the sphere, $\beta$ and $\gamma$ are
normalization constants. We have added to the $d$-dimensional action the
cosmological term (by $R \mapsto R + 2\Lambda$) and included into $V$
the electromagnetic term coming from
$X = -\EXP^{-2\phi}/\beta$. As was pointed
out in \cite{Soda}, the parameter $\nu$ may formally assume any real value.
In particular,
for $\nu = 0$ and $Q^2 = \gamma = 0$ we obtain the CGHS dilaton gravity.
However, to get the CGHS model coupled to the scalar field we have to change
$Z$ to $Z_0 = -\gamma_0$ (and choose $y(\psi) = 0$).

Many other models (describing black holes, strings, cosmologies)
can be written in the form (1). The approach of our paper may also
be used for coupling of DG to many scalar fields,
nonabelian gauge fields, spinor fields, etc. To compare different models
considered in a rather extensive literature on this subject, one has
to keep in mind that,
classically, different parametrizations of the potentials in (1) and
the Weyl transformations of (1) can be used. For example, for positive
definite $U$ we may use the representation $U \equiv \EXP^{-2\phi}$ or
$U \equiv \phi^2$. For simplicity, we will often use $U \equiv \phi$.
In classical theory, it is always possible. In quantum theory,
this parametrization is not necessarily equivalent to the exponential
one or to $U \equiv \phi^2$, etc. (see e.g \cite{Banks}).
The Weyl transformation $\metl = \Omega (\phi) \metlb$
is even more dangerous. However, in the
classical framework, it can also be used to compare differently looking
models. If two Lagrangians can be identified by using Weyl rescaling
and a different choice for $U$ in terms of $\phi$,
they are equivalent.

To find the Weyl transformation of (1) we only need to know that
$R$ transforms in the following way \cite{Wald}
$$
R = \Omega^{-1}(\bar{R} - {\bar{g}}^{ij} \nabla_i \nabla_j \log{\Omega}),
$$
where $\bar{R}$ is the curvature in terms of $\metlb$ and $\nabla_i$
is the corresponding covariant derivative. The transformed Lagrangian
$\bar{\cal{L}}$
may now be written in the form (1) with $U$ and $Z$ unchanged and
$\bar{V} = \Omega V$, $\bar{Y} = \Omega Y$, $\bar{X} = \Omega^{-1} X$
(because ${\bar{F}}^{ij} = \Omega^2 F^{ij}$).
The transformation of $W$ is more complicated,
$$
\bar{W} = W + U^\prime (\log{\Omega})^\prime \, ,
$$
but $w$ transforms simply, $\bar{w} = \Omega w$. Note that, by definition,
$\bar{f} = \Omega^{-1} f$.
Note also that $X$ and $V$ transform differently but
$X \eml \emu$ transforms like $V$. Thus $f w \equiv \tilde{f}$,
$V/w \equiv \tilde{V}$, $X \eml \emu /w$  and $Y/w$
are  invariant under the Weyl rescaling.  Using this fact it
is easy to prove that the equations (2), (3), (5) and (6) are invariant.
We will later show that Eq.(4) is also invariant.

\medskip
\noindent {\bf 3.}
Let us first discuss integrability of the models with $Y=Z=0$.
Using (2), which now is $(U_i/\tilde{f})_i =0$, one may write (3) in two
equivalent forms (recall that the electromagnetic term is included in $V$)
$$
U_{uv} {U_i / \tilde{f}} + \tilde{V} U_i =0. \; (i = u, v),
$$
 Defining a new function
$N(U)$ by $N^{\prime}(U) = \tilde{V}$ we then see that $M \equiv
N(U) + U_u U_v /\tilde{f}$ is locally conserved, i.e. $M_u = 0$ and
$M_v =0$. We may rewrite this Weyl invariant quantity in the coordinate
invariant form
$$
M \equiv N(U) + U_u U_v /\tf = N(U) - \metu U_i U_j /w \;.\;\; \eqno(8)
$$
It is not difficult to prove that the most general locally conserved
scalar depending on $\phi$, $\phi_i$ and $\metu$
must be an arbitrary function of $M$. To show this, one may use the
equations (2) and (3) to derive $\dif_k F(\phi, \metu \phi_i \phi_j)$,
where $k = u, v$.

Using the local conservation of $M$, it is very easy to solve the equations
(2)-(4). From (2) we have
$$
\tilde{f}/U_u = b^{\prime}(v),\;\;\;\tilde{f}/U_v = a^{\prime}(u),
								\eqno(9)
$$
where $a(u)$ and $b(v)$ are arbitrary functions. This equation shows
that $U_a = U_b$ and thus $U$ (and $\phi$) depend only on one variable
$\tau \equiv a(u) + b(v)$. From (8) and from the conservation of $M$, it
now follows that $f$ can be cast into the form
$$
f(u,v) = h(\tau) a^{\prime}(u) b^{\prime}(v), \;\;\;
\tau \equiv a(u) + b(v).\;\; \eqno(10)
$$
This means that all the equations (2)-(6) are in fact ordinary differential
equations for the functions of one variable $\tau$
(in what follows we denote the derivatives in $\tau$ by dots).
Then, using (8) and (9), we express $h$
in terms of $U$ (or $\phi$), namely, $h = (M - N) w^{-1}$. Now Eq.(9)
tells us that $\dot{U} = h w$ and thus $U$ satisfies the equation
$\dot{U} = M - N(U)$ that gives $\tau$ in terms of $U$.
This completes the solution of all the equations.
The finite roots, $U = U_0$, of the equation $M = N(U)$ define apparent
horizons. Near each horizon one can introduce Kruskal-like coordinates
and then study the structure of the singularity, as has recently been
done \cite{Kummer} for a special class of DG~\footnote{
Note that the authors have found the singularity structure
consistent with the SBH for the DG models having $U$, $V$, $W$ given by
(7a) with $\Lambda = Q = 0$. This result is quite natural as these models
describe the standard spherical black holes in the $d$-dimensional
Einstein gravity.}.

As will be shown below for a more general case, the equations for the
functions of one variable $\tau$ can be written in the Hamiltonian form,
which is most suitable for quantizing in our approach \cite{VDA}. Note
that there we used a different metric, which is equivalent to the metric
used here due to a simple lemma: two metrics,
$$
ds^2 = -4 f(u,v) du dv\;,\;\;\;ds^2 = -4\alpha(r)dt^2 +
4\beta(r)dr^2,\;\;
$$
are related by a coordinate transformation $u = u(t,r),\s v= v(t,r)$
if and only if there exist two functions $a(u)$ and $b(v)$, such that
Eq.(10) is satisfied. Moreover, this transformation depends on one
arbitrary function $c(r)$, and also $\alpha(r) = - h(2c(r))$, $\beta(r) =
a(r) c^{\prime 2}(r)$. This lemma shows that the metric is static
if and only if $f$ can be represented in the form (10). Thus the
above direct derivation of (10) from the field equations reproduces
the generalized Birkhoff theorem.

Finishing with DG models I have to emphasize that the above results are
essentially not new -- I simply derived them in a most general and direct
way suitable for a generalization. The locally conserved scalar
$M$ was introduced in \cite{Frolov} and generalized in several papers
(see \cite{Mann}, \cite{Uehara} and references therein). For the
$d$-dimensional SBH, $M$ is proportional to the black hole mass.
The canonical mass of SBH introduced in our papers \cite{VDA} is
of course equivalent to (8). Various other definition of the mass
are compared in \cite{Lau}. The solution of the DG equations based on
conservation of $M$, which is the substance of the generalized Birkhoff
theorem, was constructed (in a different form) and applied to SBH physics
in \cite{Kunst}.

Before proceeding to solving DGS models, I would like to mention that
the equations of DG may be presented in a very simple form. Defining a new
function $\Phi(U)$ by $\Phi^{\prime}(U) \equiv (M - N(U))^{-1}$, it is easy
to see that Eq.(3) implies that $\Phi$ satisfies the D'Alembert equation
$\Phi_{uv} = 0$. It is not difficult to prove that, conversely, the
equations (2), (3) follow from the three equations
$$
M_u = 0,\;\;M_v = 0,\;\;\Phi_{uv} = 0,\;\; \eqno(11)
$$
with the above definitions of $M$ and $\Phi$ (by the way, Eq.(4) is
satisfied for all solutions of (2) and (3)). Though the equations (11) look
simple, it is not clear how one can directly apply them to quantizing
black holes and other DG models, while the approach of \cite{VDA},
first reducing DG to FDC, gives an explicit construction of the Hilbert
space and observables (this yet remains to be demonstrated for the
general DG). Nevertheless, a formulation based on the equations (11)
might prove to be useful for physics interpretation
and applications including, say, nonspherical or nonstatic perturbations.

\medskip
\noindent {\bf 4.}
Unfortunately, the scalar $Z$-term can not be treated as a perturbation
of the integrable DG model. Instead, we have to look for explicitly
integrable models with nonzero $Z$. Then we might hope to treat
nonintegrable terms as a perturbation (as we will see below, general
DGS models must be not integrable). We first consider a generalization of
the string-inspired models, for which $U, V, W$ are arbitrary but $Y=0$
and $Z = Z_0 = -\gamma_0$ is independent of $\phi$. As explained above,
the $X$ term may be included in $V$ and we omit it. To make the
presentation more compact, we choose $U = \phi$ and use the Weyl
invariant combinations $\tilde{f}$ and $\tilde{V}$ instead of $f$ and $V$.
Then, by using Eq.(3) we may rewrite Eq.(4) in a simpler Weyl - invariant
form (for $Z = Z_0$ the right-hand side is zero):
$$
(\log{\tilde{f}})_{uv} + \tilde{f} \tilde{V}^{\prime} = Z^{\prime}
\psi_u \psi_v \, .                                                \eqno(12)
$$

Adding to this the equation (3) multiplied by a constant $g_1$, we find
$$
F_{uv} + \epsilon \,\EXP^{F} [(\tilde{V}^{\prime} + g_1 \tilde{V})
\EXP^{-g_1 \phi}] = 0\, ,                                         \eqno(13)
$$
where $F = \log{(\tilde{f} \EXP^{g_1 \phi})}$ and $\epsilon$ is the sign of
$\tilde{f}$. Now, if we choose $\tilde{V}$ so that the expression in the
square brackets is a constant, $2 g_2$, this equation will be the famous
Liouville equation that is known to be integrable  \cite{Andrei}.
The most general potential satisfying this requirement is
$$
\tilde{V} = g_3 \EXP^{-g_1 \phi} + {g_2 \over g_1} \EXP^{g_1 \phi}\;\;
                                                                  \eqno(14)
$$
(for a general $U$ we may simply replace here $\phi$ by $U$). The general
solution of the Liouville equation is known, and we may write the
solution of Eq.(13)
$$
f w\ \EXP^{g_1 \phi} = a^{\prime} (u) b^{\prime}(v) [1 + g_2 a(u) b(v)]^{-2},
\;\; \eqno(15)
$$
where $a(u)$ and $b(v)$ are arbitrary functions and so $\epsilon$ can be
included in their definition. Using this solution and Eq.(3) with the
potential (14), we can also reduce it to the Liouville equation
$$
\Phi_{ab} + g_3 \EXP^{-2 g_1 \Phi} = 0 \, .  \eqno(16)
$$
Here $\Phi \equiv \phi + g_1^{-1}\log{(1 + g_2 a b)}$ is regarded as
a function of the new variables $a, b$. The exact general solution of
(16) gives  $\Phi$
in terms of two new arbitrary functions $A(a)$ and $B(b)$
$$
\EXP^{-2 g_1 \Phi} = A^{\prime}(a) B^{\prime}(b)
[1 - g_1 g_3 A(a) B(b)]^{-2} .                              \eqno(17)
$$
Finally, we have to show that the equations (2) and (6)
are also satisfied. In fact, these equations define $\psi(u,v)$
satisfying the D'Alembert equation $\psi_{uv} = 0$ (or, equivalently,
$\psi_{ab} = 0$). To prove this, it is sufficient to show that
$[\tilde{f}(\phi_a/\tilde{f})_a]_b = 0$. This can be checked
explicitly by using (13) and (16).

A simpler model corresponds to $g_1 = 0$. As the above expressions are
singular in this limit, we define ${\bar{g}}_3 \equiv g_3 + g_2/g_1$.
Then, in the limit $g_1 \to 0$
$$
\tV = {\bar{g}}_3 + 2g_2 \phi \, .                           \eqno(18)
$$
To solve the equations with this potential we may simply put $g_1 = 0$ in
Eq.(15). Then defining new arbitrary functions  $\bar{A}(a)$ and
$\bar{B}(b)$ by $A(a) = a + g_1\bar{A}(a)$ and $B(b) = b + g_1\bar{B}(b)$
we find, by collecting the first-order terms in (17),
$$
\phi = -{{\bar{g}}_3 \over 2 g_2} - {\bar{A}}^{\prime}(a) -
{\bar{B}}^{\prime}(b)
       + 2g_2{\bar{A}(a) b + \bar{B}(b) a \over 1 + g_2 a b} \; . \eqno(19)
$$
In the limit $g_1 \to 0$ and $g_2 \to 0$, our model degenerates to
the models related to that of \cite{CGHS} by Weyl rescalings and
parametrizations\footnote{
In the classical framework, they all are equivalent and thus explicitly
integrable. However, in general, their quantum versions are not equivalent
due to anomalies
(see e.g. \cite{Jackiw}).} of $U$ (with $U = \phi$, this model is defined
by $w = \phi$, $\tV = 4\lambda^2$ and $Z = -\gamma_0$). Its solution can be
obtained by an obvious further redefinition of $\bar{A}$ and $\bar{B}$.

\medskip
\noindent {\bf 5.}
Though all the above equations are integrable and their general
classical solutions have been written
explicitly, the full quantum treatment of our integrable system is
a separate and yet unsolved problem. The exact solutions of the
s-integrable systems introduced below are easier to quantize. In addition,
they give us exact spherically symmetric static solutions of
the Einstein gravity non trivially
coupled to matter in any space-time dimension that may be useful in various
applications.

As we  have seen, even the simplest DGS models have nonstatic solutions
(we call `static' the solutions effectively depending of one variable
$\tau$). Nevertheless, the static solutions form an interesting subclass
of the solutions to Eqs.(2)-(6) if we use Eq.(10) as {\it Ansatz} and
suppose that all other fields depend on one variable $\tau$. Then the
metric is static, and we will show that the functions depending on $\tau$
are coordinates of a constrained Hamiltonian system (the constraint is the
Hamiltonian $H$ itself, i.e. $H = 0$). This system is integrable if there
exist two more integrals depending on the coordinates $h$, $\phi$, $\psi$
and velocities $\hd$, $\fid$, $\psid$ (or momenta introduced below)
and the system of the first-order
equations defined by these integrals is explicitly integrable (recall that
Eq.(5) is always integrable). Then we call the DGS model s-integrable.
If $Y = 0$, as we assume in this paper,
the equation (6) that now is simply $(Z \psid \sqd = 0$, gives the integral
$C_0 = Z \psid$. Thus the problem is to find one more integral.

The remaining ordinary differential equations are ($U \equiv \phi$):
$$
h w (\fid /h w\sqd \equiv
\fidd - W \fids - \fid F = Z \psids \, ;                    \eqno(2^*)
$$
$$
\fidd + h V = 0\, ;                                         \eqno(3^*)
$$
$$
2W \fidd + W^{\prime} \fids + \dot{F} +
h V^{\prime} = Z^{\prime} \psids  \, .                       \eqno(4^*)
$$
For convenience, we will use the notation $F \equiv \dot{h} / h$
(not to be confused with $F$ above).
From $(2^*)$ and $(3^*)$ we immediately derive the integral
$$
L \equiv W \fids + \fid F  + h V + Z \psids = 0 .             \eqno(20)
$$
We will see shortly that $L$ is the Hamiltonian of a FDC system giving
all the equations of motion if it is constrained to be zero.

To derive an additional integral, we exclude $\psids$ from $(4^*)$ by
using $(2^*)$
$$
(2 W Z - Z^{\prime} ) \fidd + (W Z)^{\prime} \fids +
h V^{\prime} Z + (F Z\sqd = 0\, .                             \eqno(21)
$$
Here we may either replace $h$ by $-\fidd /V$ or exclude $\fids$ by
using (20). In both cases we can find potentials for which (21)
has an integral.

The first approach allows us to rewrite (21) in the form
$$
(A(\phi)\fid \sqd + (Z F \sqd = 0\s , \;\;\;A(\phi) \equiv 2 Z W -
Z^{\prime} - Z {V^{\prime} / V} \s ,
$$
if $A^{\prime}(\phi) = (Z W)^{\prime}$. This gives the differential
equation $[\log{(\tV Z)}]^{\prime} = g_1 Z^{-1}$, where $g_1$ is an
arbitrary integration constant. Solving this equation
we may express $Z$ in terms of $\tV \equiv V/w$, or $\tV$ in terms of $Z$
(recall that $N^{\prime}(\phi) \equiv \tV$):
$$
Z = (g_2 + g_1 N(\phi)) {\tV}^{-1} \, ,\;\;\;\tV =
g_2 \EXP^{g_1 z(\phi)} Z^{-1} \, ,                           \eqno(22)
$$
where $z^{\prime}(\phi) \equiv Z^{-1}$.
With the potential satisfying (22), we have the integral
$$
Z F + (Z W - g_1) \fid = C_1\s .                            \eqno(23)
$$

The three available integrals $C_0$, $C_1$ and $L$ allow us to
find the general solution to all the equations. Indeed, using the above
equations, it is not difficult to express $\fid$ in terms of
$C_0$, $C_1$ and of $x \equiv h V Z$. Then one can derive
the following first - order equations for $\phi(\tau)$ and $x(\tau)$
$$
2 g_1 \fid = -C_1 + R(x)\, ,\;\;\;\;\dot{x} = x R(x)/Z(\phi)\, ,  \eqno(24)
$$
where we introduced the notation
$R = [C_1^2 - 4 g_1 (C_0^2 + x)]^{\half}$
(note that the square root may be positive or negative).
From these two equations we derive one explicitly integrable
equation
$$
2 g_1 x {dz \over dx} = 1 - C_1 R^{-1}(x)\, .                     \eqno(25)
$$
The solution of this equation expresses $z$ as an elementary function
of $x$, $C_0$, $C_2$ and of one additional integration constant (let me
omit this simple but tedious algebra). To complete integrating the whole
system, we have to express $\phi$ in terms of $z$  using the
definition of $z(\phi)$. Then $Z(\phi)$ can be expressed in terms of
$z$, which is a known elementary function of $x$, and the second equation
in (24) is reduced to a quadrature (in general, not elementary).

The second approach to deriving an additional integral
is to exclude from (21) the terms containing the
derivatives of $\phi$ by using $(3^*)$ and (20). In this way we obtain
the equation (we also replace $h$ by the identical expression
$\dot{h} / F$)
$$
F [(F Z\sqd - F Z (\log{W Z}\sqd \s ] + \dot{h} V Z [(\log{V/W})^{\prime}
- 2W] - C_0^2 (\log{W Z})^{\prime} (\log{h}\sqd = 0.
$$
Dividing this by $\half W^2 Z$ we make the first term the total derivative
$(F^2/W^2\sqd$. Now, to obtain an integral we choose the
potential so as to make the multipliers of $\hd$ and $\Lhd$ constant.
Solving the corresponding equations for the potentials, we derive
the most general potentials for which the above equation is the
total derivative of
$$
(F/W)^2 + 4 {\bar{g}}_1 h + 2 {\bar{g}}_2 C_0^2 \Lh = {\bar{C}}_1\ ,
                                                                  \eqno(26)
$$
which is the integral we are looking for. The potentials satisfying our
requirement are now defined by two relations (it must be clear now why it
was important to keep $w$ arbitrary):
$$
V = W({\bar{g}}_4 w^2 - {\bar{g}}_1),\;\;\; Z^{-1} = W ({\bar{g}}_3 +
{\bar{g}_2} \log{w}),                                            \eqno(27)
$$
where ${\bar{g}_1}$ - ${\bar{g}_4}$ are arbitrary real constants.
Using (26) and (20) one can derive the first-order differential equations
for $h$ and $\bar{h} \equiv h w^2$
$$
\dot{\bar{h}} = \bar{h} W R_1(\bar{h})\, , \;\;\;\; \hd = h W R_2(h) \, ,
							         \eqno(28)
$$
where
$$
R_1 = ({\bar{C}}_1 - 4{\bar{g}}_3  C_0^2 - 4 {\bar{g}}_4 \bar{h} -
2{\bar{g}}_2 C_0^2 \log{\bar{h}})^{\half} ,
\;\;R_2 = ({\bar{C}}_1 -4{\bar{g}}_1 h - 2{\bar{g}}_2 C_0^2 \log{h})^{\half}.
$$
Thus, to find the relation between $\phi$ and $h$ we have to solve the
equation
$$
{d \bar{h} \over dh} = {\bar{h} R_1 (\bar{h}) \over {h R_2(h)}} \ .
                                                                  \eqno(29)
$$
Integrating this equation reduces to one non elementary quadrature (it
is elementary if ${\bar{g}}_2 = 0$).

By the way, with ${\bar{g}}_2 = 0$ we can derive the exact static spherically
symmetric solution of the Einstein gravity
coupled to the scalar and electromagnetic fields in any space-time
dimension (not necessarily integer). The potentials in Eqs.(7) satisfy
Eq.(25) if (recall that we must choose the representation $U = \phi$,
in which $W = (1 - \nu)/\phi$, $w = \phi^{1-\nu}$, $Z = -\gamma \phi$)
$$
2\alpha = {\bar{g}}_4 (1 - \nu), \;\;\; 2\beta Q^2 = {\bar{g}}_1 (1 - \nu), \;
\gamma^{-1} = - {\bar{g}}_3 (1 - \nu),\;\;\;\Lambda = 0.  \eqno(30)
$$
The integral of motion ${\bar{C}}_1$ is simply
$$
{\bar{C}}_1 = [F^2 \phi^2 + 8\beta (1 - \nu) Q^2 h] (1 -\nu)^{-2}.
$$
One can see that $\nu = 1$ corresponding to the space-time dimension
$d = 3$ is not allowed (this is also true for DG);
our method is applicable to any other $\nu$.

We will not consider the solutions and their physics applications in more
detail. This will be done in a forthcoming publication. Our
aim was to construct a general framework for applications which may
be pursued in different directions. However, I would like to mention
two important points. First, the solutions of the equations (25) or (28)
expressing $h$ in terms of $\phi$ have several branches depending on
the integrals $C$. With changing $C$, new solutions emerge by bifurcations.
Second, examining exact solutions
of the static equations with the scalar coupling, I have checked for
them the `no hair' theorem. Indeed, all exact solutions do not give
a black hole type singularity in the following strong sense:
for any fixed values of $C_0$, $C_1$ (or ${\bar{C}}_1$) and any finite
value of $\phi$ the metric function $h$ has no zeroes. In fact, to
prove that $h$ has no zeroes, it is sufficient to use Eq.$(3^*)$
and the constraint (20) that are  valid in general, without integrability
(under rather mild restrictions on the potentials and using some local
existence theorem for nonlinear differential equations). One may also
relax the condition $Y = 0$, consider more general coupling to several
scalar fields, etc.
Although it would be not very difficult to prove the above statements,
this would distract us from the main subject of this paper and these
results will be presented in a separate publication.

\medskip
\noindent {\bf 6.}
The final subject of this paper is the Hamiltonian formulation of
the `static' equations of DGS. It is not difficult to show that
Eqs.$(2^*) - (4^*)$ as well as the constraint (20) and the omitted
equation $(Z \psid \sqd = 0$ can be derived by varying the Lagrangian
$$
{\cal L}^{(s)} \equiv (L - hV)/l - lhV                         \eqno(31)
$$ in all variables including the Lagrangian multiplier $l(\tau)$
(Eq.(20) is reproduced in the gauge $l = 1$).
Introducing the momenta
$$
lp_{h} = \fid/h\, , \;\;lp_{\phi} = 2W \fid + \hd /h\, , \;\;lp_{\psi} =
2Z \psid \, ,
$$
we find the Hamiltonian ${\cal H}^{(s)} = l H^{(s)}$, where
$$
H^{(s)} = h p_h p_{\phi} - W h^2 p_h^2 + h V + p_{\psi}^2/4 Z \, .
                                                              \eqno(32)
$$
Now one may express the integrals of motion in terms of the canonical
coordinates and momenta and check that their Poisson brackets with
$H^{(s)}$ vanish when these variables satisfy the canonical equations
of motion (including the constraint $H^{(s)} = 0$). Of course, the
canonical equations are equivalent to the Lagrangian ones that were solved
above (with $l = 1$)\footnote{The necessity of this gauge fixing is
related to the fact that we were using the conformally flat metric
depending on one function $f(u,v)$.
If we would use the metric defined by $\alpha(r)$ and $\beta(r)$, a
Lagrangian multiplier will emerge automatically \cite{VDA}.}.

Note that the form of the Lagrangian and Hamiltonian is not
uniquely defined due to a freedom in the choice of the Lagrangian
multiplier $l(\tau)$. We may multiply $\ham$ by a function
of the coordinates, $\lambda(h, \phi, \psi)$ (not having zeroes inside
the domain of definition of the coordinates) and correspondingly divide
$l(\tau)$ by $\lambda$. The new Hamiltonian
${\Hamb} \equiv \bar{l} {\hamb}$ define the same equations of
motion due to the constraint $\ham = 0$. Such a freedom
is useful because the new Hamiltonian may have additional integrals of
motion. This observation was used in our approach to quantizing
black holes \cite{VDA}. In that case, the mass of the black hole
is proportional to $M$ defined in (8). It is conserved
when the scalar field is completely decoupled, i.e. for
$C_0 = 0$.
For the `static' case, the mass function is simply
$$
M = N(\phi) + h p_h^2 /w \equiv N(\phi) + \fids /h w \ .  \eqno(33)
$$
When $p_\psi \equiv 2 C_0 \neq 0$, $M$ is not conserved but, for s-integrable
models there may exist other integrals of motion \cite{Marco}. In the
s-integrable models, the integrals $C_1$ and ${\bar{C}}_1$ found in this
paper play the role of $M$. Thus, it is not difficult to show that
$$
C_1 = -g_1 {w \over p_h} \left( M + {g_2 \over g_1} +
{p_\psi^2 \over 4 g_1 h w} \right).                        \eqno(34)
$$
When $p_\psi \equiv 2 C_0 = 0$ the factor $w/p_f$ in (34) becomes an
additional integral of motion because $C_2 \equiv p_h /w = \fid /h w$ is
independent of $\tau$ due to Eq.$(2^*)$. A relation similar to (34)
may be derived for ${\bar{C}}_1$. We write it only for $p_\psi = 0$:
$$
{\bar{C}}_1 = (M^2 - 4 {\bar{g}}_1 {\bar{g}}_4)/ C_2^2 \ .
$$

Of course, the Lagrangian and Hamiltonian formulations are valid
in general, when additional integrals of motion are not known or
even do not exist. For nonlinear Hamiltonian systems with two or more
independent coordinates the existence of such an integral is a rare
event. Apparently simple systems with two coordinates are not
integrable and thus exhibit complex phenomena known as dynamical
chaos. A famous example is the Henon - Heiles system of two oscillators
with cubic couplings between them (see e.g. \cite{chaos}). If one
compares our general `static' system (32) with fixed $p_\psi$ to such
well-studied nonitegrable systems, one will hardly believe that it can
be integrable (integrability of the DGS field theory looks even less
probable). This does not mean that further examples of integrable DGS
can not be discovered. Of course, our simple approach
is not a suitable framework
for such a general search while general criteria for integrability
are not known. Yet, one may try to compare DGS equations to known
classes of integrable systems \cite{Andrei}, \cite{Asik}. Especially
interesting are systems related to solitons \cite{Ludwig}, \cite{Serge}.
On the other hand, chaotic phenomena in classical
nonintegrable DGS models might be of significant physics interest.
In quantum framework, nonintegrable models may still be useful if
they can be treated perturbatively on some explicitly integrable
background (like our s-integrable models).

\medskip
\noindent {\bf 7.}
In summary, we have studied a general dilaton gravity coupled to
electromagnetic and scalar fields described by Eq.(1). For the
$U - X$ models (DG) we presented some known results in a more compact
and, hopefully, simpler form (e.g. Eq.(11)). For the $U-Z$ models
(DGS) we concentrated on a somewhat simpler case $Y = 0$ and obtained
a class of integrable theories with the constant potential $Z$, thus
generalizing the well-known CGHS dilaton gravity (equations (13) - (19)).
For DGS with arbitrary $Z$ we have found two classes of s-integrable
systems and constructed for them exact `static' solutions (equations
(21)-(29)).
A special case of the s-integrable DGS gives new exact solutions of the
Einstein gravity coupled to matter in any space-time dimension. We also
pointed out a certain generalization of the `no-hair' theorem.
By constructing the Hamiltonian formulation of the s-integrable systems
we have paved a way to their quantizing. A detailed discussion of these
results and further applications will be presented elsewhere.

\section*{Acknowledgments}
This paper has been started and completed while the author was visiting the
Research Institute for Mathematical Sciences of the Kyoto University.
It is a pleasure to thank Noboru Nakanishi for organizing this visit,
kind hospitality and useful discussions. I am also grateful to participants
of his seminar on quantum gravity for interest in this work. For invariable
support and stimulating discussions I am grateful to Kazuhiko
Nishijima and Vittorio de Alfaro. Some results were discussed with
Takehisa Fujita, Vladimir Korepin, Dieter Maison, and Evgenii Sklyanin
who made useful remarks. This investigation was partially supported
by the Russian Fundamental Science Foundation (project 95-02-05679),
and by INTAS (project 93-0127).

\end{document}